\documentclass[11pt,dvips]{article}
\usepackage{epsfig,times} 
\usepackage{picinpar}
\usepackage{wrapfig}
\usepackage{floatflt}
\makeatletter
\renewcommand{\section}{\@startsection{section}{1}{0in}
	{0.4\baselineskip}{0.1\baselineskip}{\Large\bf}}
\renewcommand{\subsection}{\@startsection{subsection}{2}{0in}
	{0.25\baselineskip}{-\baselineskip}{\large\bf}}
\renewcommand{\subsubsection}{\@startsection{subsubsection}{3}{0in}
	{0.1\baselineskip}{-\baselineskip}{\normalsize\bf}}
\makeatother
\newcommand{\icrc}{$26^{\rm th}$ ICRC\ }
\begin{document}

%
\thispagestyle{myheadings}
%
\markright{OG 4.3.35}
\begin{center}
%
{\LARGE \bf VERITAS: Performance characteristics (baseline design).}
\end{center}

\begin{center}
%
%
{\bf V.V. Vassiliev$^{1}$, D.A. Carter-Lewis$^{5}$, A.M. Hillas$^{2}$, 
M.P. Kertzman$^{3}$, J. Knapp$^{2}$, \\ F. Krennrich$^{5}$, 
R.W. Lessard$^{4}$, H.J. Rose$^{2}$, G.H. Sembroski$^{4}$}\\

{\it $^{1}$ FLWO, Harvard-Smithsonian CfA, P.O. Box 97, Amado, AZ 85645, USA \\
$^{2}$ University of Leeds, Leeds LS2 9JT, UK \\
$^{3}$ De Pauw University, Greencastle, IN 46135, USA \\
$^{4}$ Purdue University, West Lafayette, IN 47907, USA \\
$^{5}$ Iowa State University, Ames, IA 50011, USA}
\end{center}

\begin{center}
{\large \bf Abstract\\}
\end{center}
\vspace{-0.5ex}
%
%

VERITAS is a proposed major ground-based gamma-ray observatory 
to be built at the Whipple Observatory in southern Arizona, USA. 
It will consist of an array of seven 10m imaging Cherenkov telescopes 
designed to conduct gamma-ray observations in the energy range 
of 50 GeV - 50 TeV. A description of the baseline VERITAS design 
and optimization criteria are presented. We provide basic characteristics 
of the array performance for observations of point sources, 
such as angular resolution, energy threshold, energy resolution, 
and integral flux sensitivity. The limiting factors of the VERITAS 
performance are discussed.

%

\vspace{1ex}

%
%
\section{Introduction:}
\label{intro.sec}

Recent discoveries in ground-based Very High Energy (VHE) astronomy (reviewed 
in Ong 1998) has been achieved due to two major advances in the atmospheric 
Cherenkov technique; imaging (Hillas 1985, Fegan 1997), and stereoscopy 
(Aharonian et al. 1997a, 1997b, Krennrich et al. 1995) of the observations 
of atmospheric cascades. The former, pioneered at the Whipple and Crimean
$\gamma$-ray observatories, has been adopted now by most existing 
ground based $\gamma$-ray instruments. The latter, demonstrated
by the HEGRA collaboration, is now being considered as a prime 
technique for the next generation of ground-based VHE observatories:
VERITAS (Weekes et al. 1999), HESS (Aharonian et al. 1999), and 
NEW CANGAROO (Tanimori et al. 1999). The scientific goals of these 
projects are described elsewhere (Weekes et al. 1999), the summary of 
the physics highlights to be accomplished with VERITAS instrument are 
presented by Bradbury et al. (1999). In this submission, we summarize 
the technical characteristics of VERITAS and discuss the major performance 
parameters of this proposed VHE observatory.

\section{Technical characteristics:}
\label{tech.sec}

\begin{figwindow}[1,r,%
{\mbox{\epsfig{file=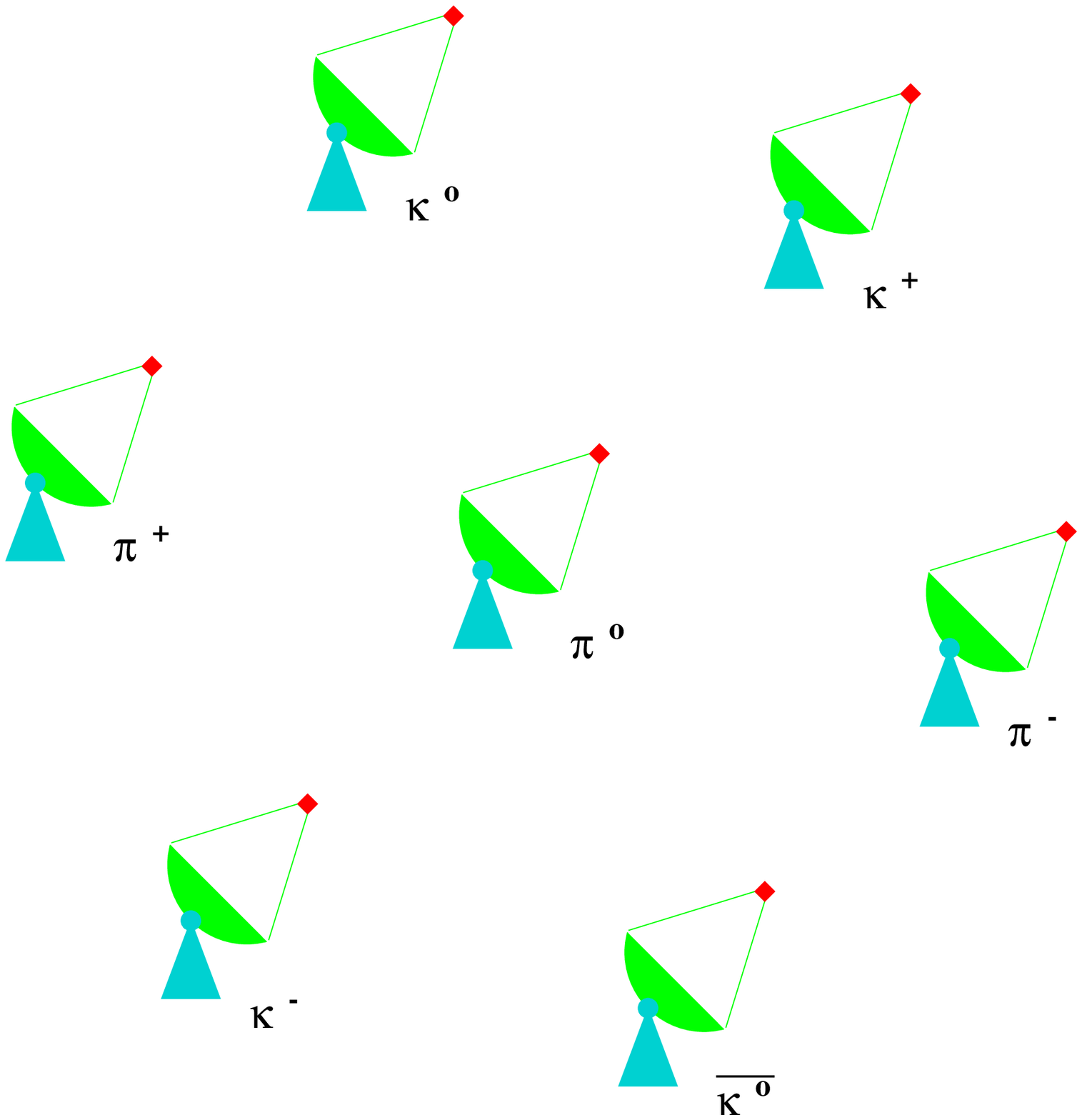,width=2.4in}}},%
\label{layout}
{{\small VERITAS hexagonal layout}}]
The VERITAS design has been optimized for maximum sensitivity 
to point sources in the energy range $100$ GeV - $10$ TeV,
but with significant sensitivity in the range $50$ GeV - $100$ GeV
and from $10$ TeV to $50$ TeV. Optimization has been performed 
with fixed total number of channels which determines the cost 
of the project. The suggested layout of the array is shown in 
Figure~\ref{layout}, and its specifications are provided in 
Table 1. Brief arguments for the baseline 
configuration can be found in Vassiliev et al. (1999), while design 
simulations of VERITAS are described in Weekes et al. (1999).
The underlying motivations for the array design were derived
from the physics goals of the project to create an instrument 
sensitive to $\sim 100$ GeV photons with high angular 
and energy resolutions, but also versatile enough 
to accomplish a variety of astronomical tasks: point source 
observations with a low energy threshold, observation of 
extended sources, sky surveys, and simultaneous monitoring 
of several objects. Some of the physics goals would require 
a different, sometimes incompatible, optimum VERITAS design. $\ $For example,
\end{figwindow}
\newpage
\begin{minipage}[t]{0.59\linewidth}
\begin{center}
Table 1: Specifications of the baseline VERITAS design.
\end{center}
\begin{center}
\begin{tabular}{r|l} \hline\hline
{\bf Location}& Montosa Canyon, Arizona, USA \\
{\bf Array elevation}& $1390$ m  a.s.l. \\
{\bf Number of telescopes}& $7$ (hexagonal layout)\\
{\bf Telescope spacing}& $80$ m  \\
{\bf Mirror}& Davies-Cotton \\
{\bf Reflector aperture/area}& $10$ m / $78.6$ m$^2$\\
{\bf Focal length}& $12$ m \\
{\bf Facets }& $244$, $61$ cm hexagon \\
{\bf Camera }& Homogeneous \\
{\bf Field of View }& $3.5$ deg \\
{\bf Number of pixels }& $499$   \\
{\bf Pixel Spacing}& $0.148$ deg \\
\end{tabular}
\end{center}
\end{minipage}
\begin{minipage}[t]{0.376\linewidth}
\noindent
detection of high energy photons ($10-50$ TeV) and
observation of objects with large angular extent ($>1^{\circ}$) 
generally require a larger field of view for the instrument.  
VERITAS will accomplish such tasks by observation of astrophysical 
objects at large zenith angles and by different array operation 
modes, e.g., offset pointing of individual telescopes to cover 
an extended object. At the same time the small, $0.15^{\circ}$, pixel size 
of the telescope cameras provides a substantial sensitivity to photons
with energies below $75$ GeV, the expected energy threshold of VERITAS 
for point source observations. Approximately $\ $20\% of the photons, which 
\end{minipage}
are detected and successfully reconstructed, will
have energies lower than $75$ GeV extending the sensitive energy 
range of VERITAS to at least $50$ GeV.

\section{Performance characteristics:}
\label{performance.sec}

\begin{figwindow}[1,r,%
{\mbox{\epsfig{file=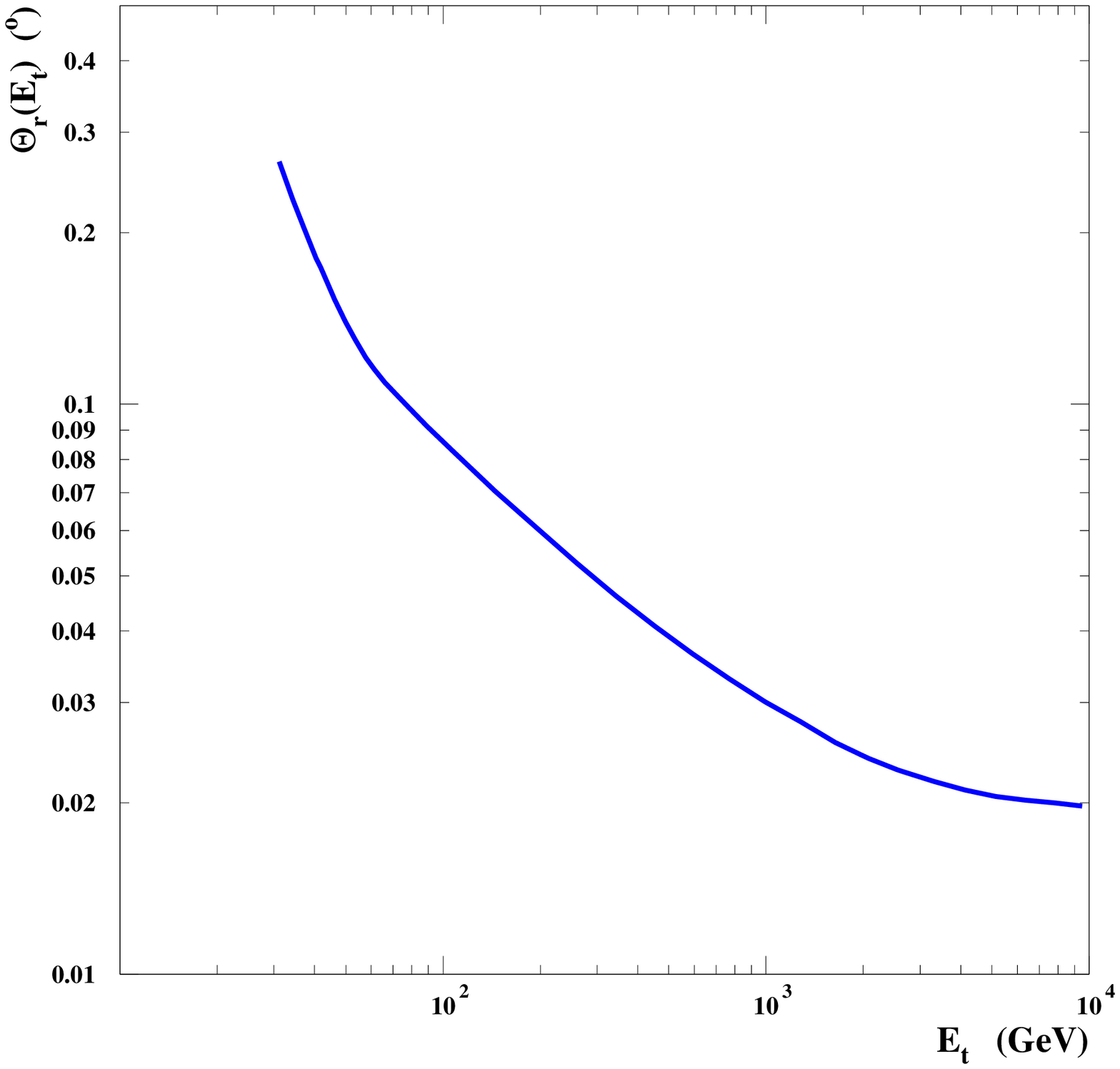,width=3.0in}}},%
\label{aresolution}
{{\small Angular resolution of VERITAS for a single photon as a function of
array energy threshold. Selection criteria for photons correspond to
VERITAS operation with maximum sensitivity to a point source.}}]
The energy threshold of VERITAS is limited by fluctuations of the Night Sky 
Background (NSB). To suppress spurious accidental signals,
a pattern trigger has been developed and tested (Bradbury et al. 1999). 
To operate telescopes at the highest rate and minimum energy 
threshold, a $500$ MHz flash ADC system will be used (Buckley et al. 1999)
which virtually eliminates the dead time of the array. Utilizing these
technologies, we expect that VERITAS will be able to operate at a
threshold of $4-7$ photoelectrons (pe) per pixel requiring coincidence 
between 2,3 telescopes of the array within $40$ nsec, and coincidence between
2,3 adjacent pixels of the cameras within $15$ nsec. Depending on the 
brightness of the different regions of the sky we expect to achieve
an energy threshold of $70-100$ GeV for point source 
observations with maximum sensitivity. The energy threshold, $E_t$, of 
VERITAS is defined here as the photon energy at which the differential 
detection rate of the photons (retained for analysis after all selection cuts)
from a source with spectrum $\propto E^{-2.5}$ is maximal. Thus, the array 
energy threshold is directly related to an array trigger threshold defined
by a hardware or software cut on the number of photoelectrons in the second 
or third adjacent pixel of the shower image. 
\end{figwindow}

When the array operates at the lowest energy threshold, requiring a trigger of 
3 adjacent pixels and 3 out of 7 telescopes, the collection area of VERITAS 
will be $1.1-7.4\times 10^4$m$^2$ for $100$ GeV, 
$10-25 \times 10^4$m$^2$ for $1$ TeV, and  $13-34\times 10^4$m$^2$
for $10$ TeV. The upper bounds correspond to all photons from the point
source which trigger the array and whose arrival direction is reconstructed
within the camera field of view. The lower bounds correspond to photons
which satisfy strict reconstruction criteria which effectively remove 
the cosmic-ray background allowing observations with maximum sensitivity.
The high angular resolution of the array (shown in Fig.~\ref{aresolution})
is one of the most important characteristics which determines the sensitivity 
of VERITAS to point sources. We expect that VERITAS will have 
better angular resolution than any existing detector operating above
a few MeV. The excellent angular resolution of the array is due to 
stereoscopic imaging. For a single telescope, the photon arrival direction 
is better defined in the direction perpendicular to the main axis of the 
image. The parallel direction is reasonably constrained by image 
ellipticity (Buckley et al. 1998). Multiple sampling of the shower from
several telescopes allows precise reconstruction of the photon origin 
in both directions. This feature of the array will also be critical 
for mapping the emission regions of extended sources with accuracy
close to one arcminute.

The performance of VERITAS is summarized by its flux sensitivity.
The minimum detectable flux of $\gamma$-rays is defined by the confidence 
level required for detection or the statistics of the detected photons. 
We require a 5$\sigma$ excess of $\gamma$-rays above the background, 
or 10 photons (below this, Poisson statistics must be used to derive
the confidence level). We estimate the flux sensitivity for 50
hours of observations on an object with a spectrum 
$\propto$ E$^{-2.5}$, which is close to the Crab Nebula spectrum seen in 
this energy range. 

\begin{figwindow}[1,r,%
{\mbox{\epsfig{file=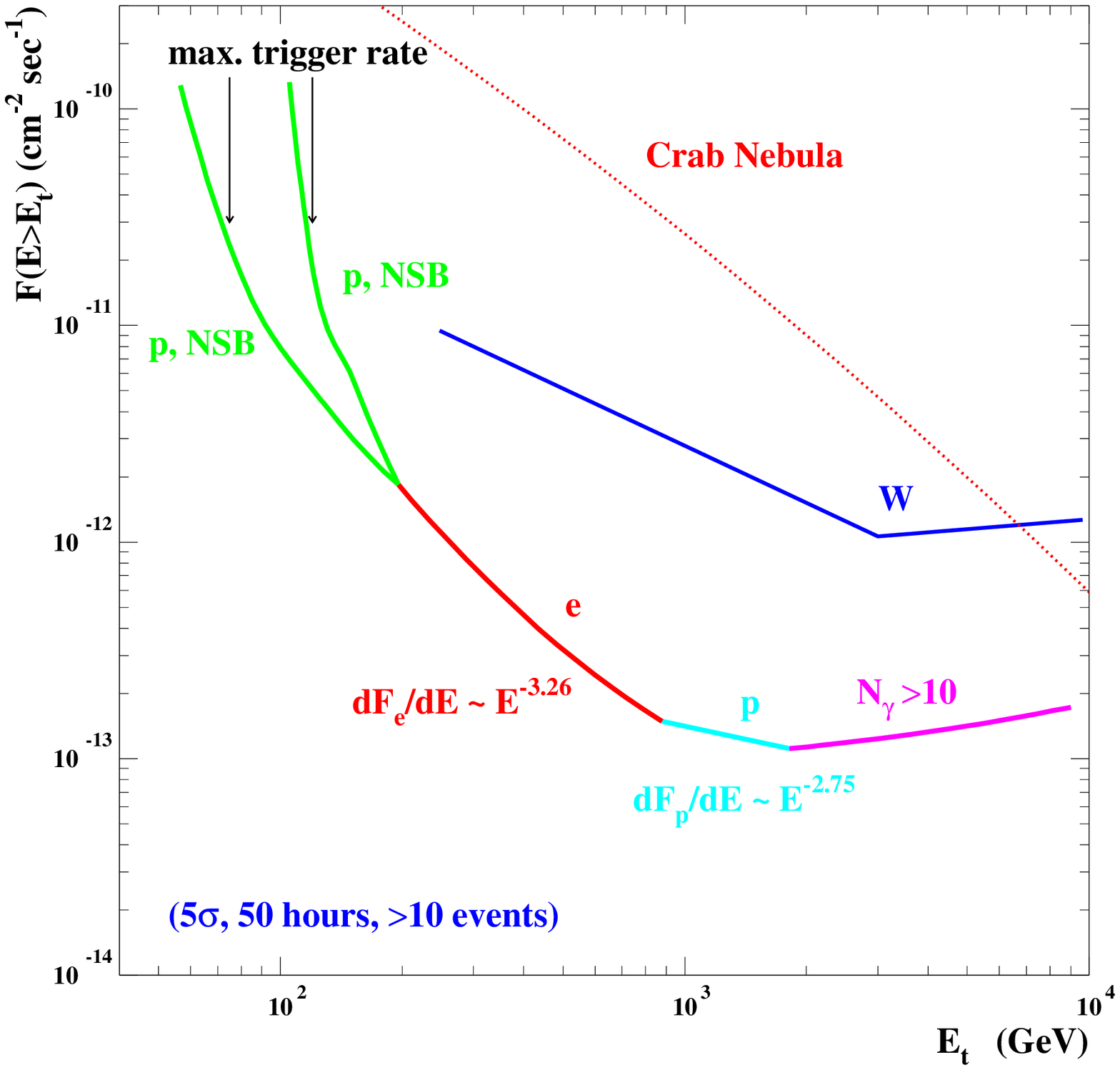,width=3.0in}}},%
\label{sensitivity}
{{\small The sensitivity of VERITAS to point-like sources in 50
hours of observing.  The dominant background as a function of energy
threshold is indicated (see text for details).  The two curves at low
energies indicate the sensitivity of VERITAS in dark (lower curve) and
bright (upper curve) NSB regions. The integral flux from the Crab 
Nebulae (Hillas et al. 1998) and estimated sensitivity of the Whipple
telescope are given for comparison.}}]
The $\gamma$-ray flux sensitivity of VERITAS for point sources as a
function of array energy threshold is shown in Figure~\ref{sensitivity}.
The complex shape of the sensitivity curve is caused by different
energy regions being dominated by the different backgrounds as
indicated in the figure. For energies above 2 -- 3\,TeV, 
the sensitivity of VERITAS is limited by photon 
statistics.  Larger telescope fields of view can improve this sensitivity 
in the future, as can large zenith angle observations. In the region 
near 1\,TeV, the sensitivity is limited by rare cosmic-ray protons 
which mimic $\gamma$-rays by converting most of their energy into 
an electromagnetic cascade in the first few interactions. A chain of 
very rare coincidences must occur for such events to pass all selection
criteria: almost all of the proton's energy must be transferred to an
electromagnetic cascade leaving no hadronic shower core; the
transverse momentum distribution of secondary photons must be very
narrow to generate a compact cascade; the axis of the shower must be
precisely aligned with the telescope axis; and the impact parameter of
such a shower cannot be large if it is to produce a well-defined
image. The rate of such events is not known exactly due to  
large variations in MC predictions caused by uncertainties in 
the different hadronic interaction models used. Therefore, we show the
estimated rate of such events. In the energy region between 200\,GeV 
and $\sim$1\,TeV, the background rejection of VERITAS is so good that 
diffuse cosmic-ray electrons are the dominant background instead of 
hadronic cosmic rays.  The diffuse electron spectrum is very steep, 
so the decrease in the sensitivity of VERITAS with decreasing energy 
is more rapid in this region.  Because electrons and $\gamma$-rays produce 
nearly identical electromagnetic cascades in the atmosphere, the only 
way to reduce this background is with improved angular resolution 
algorithms. Increasing the quantum efficiency of the photodetectors and
decreasing the pixel size in future VERITAS upgrades will also improve 
array performance in this region. The region below 200 GeV
is limited by the NSB and cosmic-ray protons.  At this level, the amount of 
collected Cherenkov light is so small that shower images
have very few pixels which pass the image cleaning process.  As such,
small fluctuations in the NSB significantly affect the reconstruction
of both $\gamma$-rays and protons.  Thus, the amount of NSB light 
determines the sensitivity of VERITAS in this region (and
thereby the energy threshold). The more sensitive of the two curves
indicates a relatively dark observation region (like an AGN with no
bright stars in the FoV) while the less sensitive curve indicates a
region where the NSB light is approximately 4 times brighter (like in
some regions of the Galactic plane). The arrows show the lowest
energy threshold at which VERITAS will be able to operate,
limited by the higher than $1$ MHz accidental trigger rates
of single telescopes caused by the NSB. We anticipate a 
counting rate of $20-40$ well reconstructed photons per minute 
from the Crab Nebula for VERITAS observations close to these limits.
\end{figwindow}
%
\begin{figwindow}[1,r,%
{\mbox{\epsfig{file=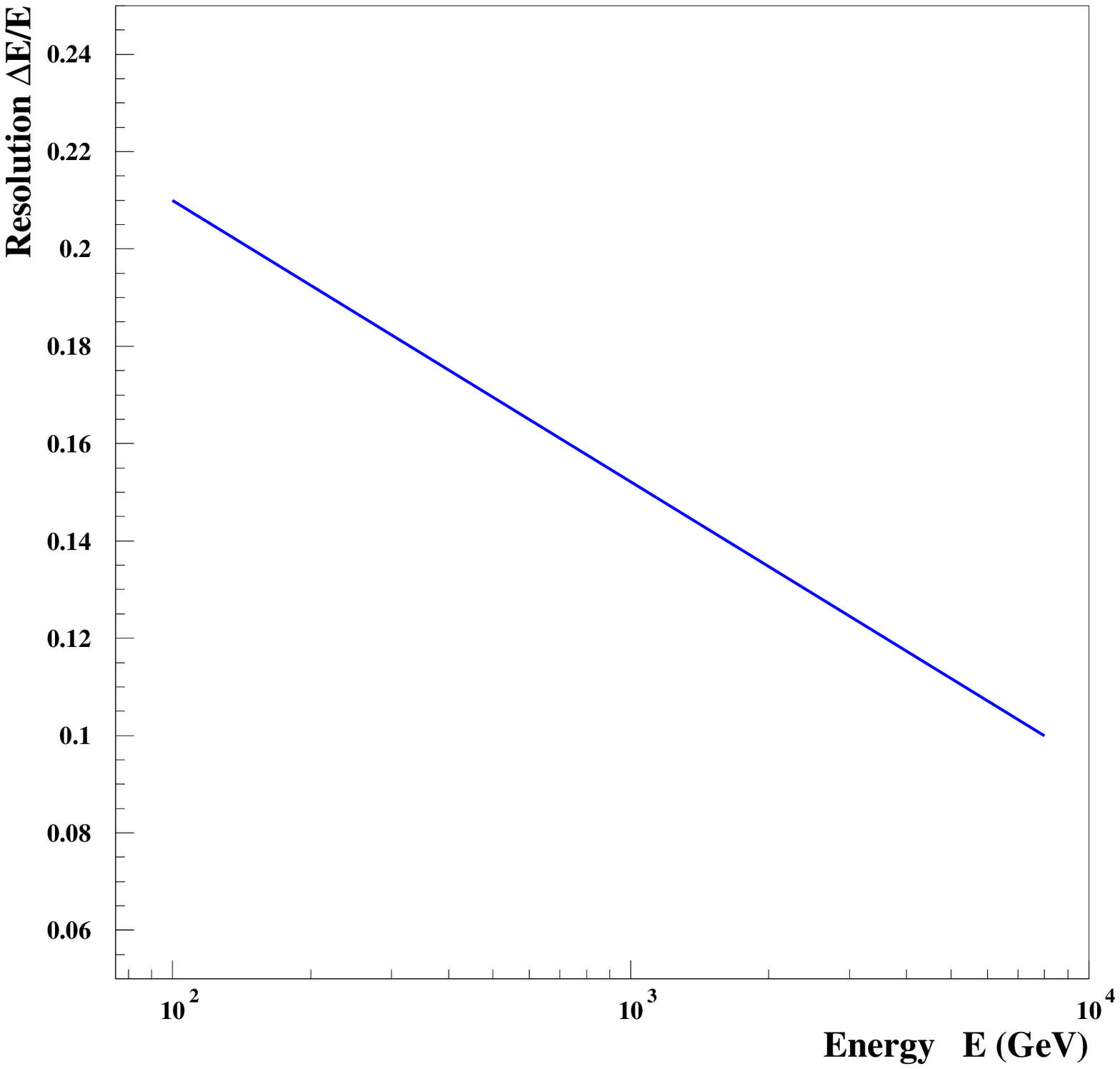,width=3.0in}}},%
\label{energy}
{{\small Estimate of VERITAS energy resolution}}]
The energy resolution of VERITAS will be considerably better than that
of the Whipple Observatory telescope for three reasons: (1) the shower
core location will be known with an accuracy of about $10$ m, (2) several 
telescopes will view each event at different distances from the shower 
core, and (3) each camera will have finer pixellation (0.15$^\circ$ vs. 
0.25$^\circ$). For the Whipple telescope, the RMS energy resolution 
using the technique described in Mohanty et al. (1998) gives 
$\Delta E/E \approx 0.35$. We adopted this method for stereoscopic
observations, as explained in Weekes et al. (1999), and obtained the
energy resolution shown in Figure~\ref{energy}. For our estimates, 
we used simulated showers from $\gamma$-rays with energies $>100$ GeV 
whose size was above $10$ pe per image, core location was 
in the range $65-180$ m from telescope, the position of the image
centroid was in the interval $0.5-1.4^{\circ}$, and whose arrival 
direction was reconstructed to within $0.1^{\circ}$ of the source 
position. The resolution improves slowly as the energy 
of the shower increases. For low energy events ($\sim 100$ GeV) the 
resolution will likely be improved through more sophisticated
energy estimates and through the use of more restrictive cuts on
events used in the energy analysis. The improved energy resolution 
of VERITAS will help resolve spectral features, 
such as a possible neutralino annihilation line from the Galactic 
center or spectral cut-offs in AGN, and permit better estimation of 
characteristics of the emission regions in sources, such as the 
magnetic field in the vicinity of SNRs and AGN.
\end{figwindow}

\vspace{0.25ex}
\begin{center}
{\Large\bf References}
\end{center}
%
Aharonian, F.A., 1997a, Astroparticle Physics, 6, 343 \\
Aharonian, F.A., 1997b, Astroparticle Physics, 6, 369 \\
Aharonian, F., et al. 1999, Proc. \icrc (Salt Lake City) OG.4.3.24\\
Bradbury, S., et al. 1999, Proc. \icrc (Salt Lake City) OG.4.3.28 \\
Bradbury, S., et al. 1999, Proc. \icrc (Salt Lake City) OG.4.3.21 \\
Buckley, J., et al. 1999, Proc. \icrc (Salt Lake City) OG.4.3.22  \\
Buckley, J., et al. 1998, A \& A, 329, 639 \\
Fegan, D.J., 1997, J. Phys. G: Nucl. Part. Phys., 23, 1013 \\
Hillas, A.M., 1985, in Proc. 19th ICRC (La Jolla), 3, 445 \\  
Hillas, A.M., et al. 1998, ApJ, 503, 744 \\
Krennrich, F., et al. 1995, Exp. Ast., 6, 285 \\
Mohanty, G., et al. 1998, Astroparticle Physics, 9, 15 \\
Ong, R.A., 1998, Physics Reports, 305, 93 \\
Tanimori, T., et al. 1999, Proc. \icrc (Salt Lake City) OG.4.3.04 \\
Vassiliev, V.V. et al. 1999, Astroparticle Physics, in press\\
Weekes, T.C., et al. 1999, VERITAS, proposal to SAGENAP           \\
\end{document}